# An Enhanced Apriori Algorithm for Discovering Frequent Patterns with Optimal Number of Scans

Sudhir Tirumalasetty[1], Aruna Jadda[2] and Sreenivasa Reddy Edara[3]

[1] Department of Computer Science & Engineering, Vasireddy Venkatadri Institute of Technology
Guntur, Andhra Pradesh 522508, India

[2] Department of Computer Science & Engineering, Vasireddy Venkatadri Institute of Technology
Guntur, Andhra Pradesh 522508, India

[3] Department of Computer Science & Engineering, Acharya Nagarjuna University
Guntur, Andhra Pradesh 522510, India

**Abstract**

Data mining is wide spreading its applications in several areas. There are different tasks in mining which provides solutions for wide variety of problems in order to discover knowledge. Among those tasks association mining plays a pivotal role for identifying frequent patterns. Among the available association mining algorithms Apriori algorithm is one of the most prevalent and dominant algorithm which is used to discover frequent patterns. This algorithm is used to discover frequent patterns from small to large databases. This paper points toward the inadequacy of the tangible Apriori algorithm of wasting time for scanning the whole transactional database for discovering association rules and proposes an enhancement on Apriori algorithm to overcome this problem. This enhancement is obtained by dropping the amount of time used in scanning the transactional database by just limiting the number of transactions while calculating the frequency of an item or item-pairs. This improved version of Apriori algorithm optimizes the time used for scanning the whole transactional database.

## 1. Introduction

Database has up-to-date vivid increase in its volume with time. This exponential increase in data resulted with an aim of discovering knowledge which is used to support decision making system. Data mining is the key step in the knowledge discovery process. The tasks of data mining are generally divided in two categories: Predictive and Descriptive. The goal of the predictive tasks is to predict the value of a particular attribute based on the values of other attributes and the goal of descriptive tasks, is to mine previously unknown and useful information from large databases. The goals of these tasks in data mining are obtained by some techniques. They are: clustering, classification, association rule mining, sequential pattern discovery and analysis. The advances of data mining systems have wide spread its magnitude in recent years for many decision making systems like sales analysis, healthcare, e-commerce, manufacturing, etc.

Among the various techniques used in discovering knowledge, association mining is one of the most central data mining's functionality. This mainly involves in extracting association rules [16]. These rules are used in identifying frequent patterns [17]. The advantages of these rules are discovering unknown relationships and generating outcomes which provides basis for decision making and prediction in areas like health care, banking, manufacturing, telecommunications etc. [16].

Existing association mining algorithms has few flaws. They are: (i) The whole database must be scanned for more number of times even though few patterns are interesting. This results in wastage of time. (ii) The rules generated by these association mining techniques are large and are difficult to understand. (iii) Defining support and confidence values is not clear. These values are defined experimentally. As a whole developing an optimal association mining algorithm is a thought provoking task.

In this paper the general Apriori algorithm and an improved version of Apriori algorithm [18] are compared which resulted in evolution of another new improved version of Apriori algorithm. This new improved version minimizes the number of database scans.

The rest of the paper is organized as follows. Section 2 elaborates about association mining. Section 3 compares the general Apriori algorithm and existing improved version of Apriori algorithm [18]. Section 4 introduces a new enhanced version of Apriori algorithm. Section 5 compares the results of existing general Apriori algorithm, existing improved versions of Apriori algorithm [18] and new improved version of Apriori algorithm. Finally, conclusion and the future scope of this new improved version of Apriori algorithm.

## 2. Association Mining

Among the existing problems in data mining association mining is predominant. Discovering frequent patterns (rules) is prevalent in association mining. These rules play a pivotal role for decision making systems and are an emerging area in research [1]. These rules provides solution to problems in areas like banking, marketing, health care, telecommunication, text databases [2], web [3] and databases containing adequate images [4].

Till dated wide number of association mining algorithms were introduced [5, 6, 7, 8, 9, 10]. These algorithms are grouped into two groups based on their approach. They are:
  i. Candidate generation approach
     Ex: Apriori [6]
  ii. Pattern growth approach
     Ex: FP Growth [9, 10]

Between these two groups, the first group developed many association mining algorithms. Among those, Apriori algorithm is the leading algorithm. This Apriori algorithm is enhanced by many scholars resulted in evolution of optimized Apriori like algorithms [11, 12, 13, 14, 15]. To discover frequent patterns these Apriori like algorithms follow iterative approach.

## 3. Existing Algorithms

3.1 The General Apriori Algorithm

The general Apriori algorithm is:

T: Transactional data base
$C_k$: Candidate item set of size k
$L_k$: Frequent item set of size k
s: Support

Apriori(T, s)
    $L_1 \leftarrow$ { large 1-item set that appear in more than or equal to s transactions }
    $k \leftarrow 2$
    While $L_{k-1} \neq \phi$
        $C_k \leftarrow$ Join($L_{k-1}$)
        For each transaction t in T
            For each candidate c in $C_k$
                If(c $\subseteq$ t) then
                    count[c]←count[c]+1
                End If
            End For
        End For
        $L_k = \phi$
        For each candidate c in $C_k$ //Prune
            If (count[c] >= s) then
                $L_k \leftarrow L_k$ U {c}
            End If
        End For
        $k \leftarrow k + 1$
    End While
    Return $L_k$
End Apriori

In general Apriori algorithm, for each candidate in $C_k$, frequency is calculated by scanning transactional database. After calculating frequencies for all candidates in a $C_k$ these frequencies are compared with support, s and exclude candidates with frequencies less than s. This results in generation of $L_k$.

The general Apriori algorithm has some flaws:
- The transactional database is scanned repeatedly. This is because every candidate of candidate set ($C_k$) generated after Join operation must be checked in all transactions of transactional database for the presence of candidate.
- If there are adequate transactions then the genera Apriori algorithm is not apt.

3.2 The Existing Enhanced Apriori Algorithm [18]

Mohammed Al-Maolegi et. al. developed an improved version of Apriori algorithm, $Apriori_1$ [18] aimed in reducing the repeated scans of transactional database. This improved algorithm $Apriori_1$ is:

T: Transactional data base
$C_k$: Candidate item set of size k
$L_k$: Frequent item set of size k
s: Support

$Apriori_1$(T, s)
    $L_1 \leftarrow$ { large 1-item set that appear in more than or equal to s transactions }
    $k \leftarrow 2$
    While $L_{k-1} \neq \phi$
        $C_k \leftarrow$ Join($L_{k-1}$)
        For each candidate c in $C_k$
            $I_x$ = Get_Item_Min_Support(c, $L_1$)
            $T_{id}$ = Get_Transaction_Ids($I_x$)
            For each transaction t in $T_{id}$
                If(c $\subseteq$ t) then
                    count[c]←count[c]+1
                End If
            End For
        End For
        $L_k = \phi$
        For each candidate c in $C_k$ //Prune
            If (count[c] >= s) then

$$L_k \leftarrow L_k \cup \{c\}$$
End If
End For
$k \leftarrow k + 1$
End While
Return $L_k$
End $Apriori_1$

In this improved version of Apriori, for each candidate c in $C_k$, item ($I_x$) with minimum support among the items in c is obtained and the transactions that contain that item ($I_x$) are grouped ($T_{id}$). Next, in each transaction t in $T_{id}$ the presence of c is checked and the frequency of c is calculated rather than scanning the entire transactional database. Later the frequencies of c in $C_k$ are compared with support, s and exclude the candidates with frequencies less than s. This results in generation of $L_k$.

The advantage of this improved Apriori algorithm is:
- The entire transactional database is not scanned for calculating the frequency of c in $C_k$.

This improved version of Apriori algorithm has a flaw:
- All transactions with transaction ids in $T_{id}$ are checked for presence of c even though few transactions contain c.

These flaws stated in section 3.1 and 3.2 are by-passed by the proposed algorithm stated in section 4.

## 4. Proposed Algorithm

4.1 The New Enhanced Apriori Algorithm

The proposed algorithm reduces number of scans when compared with $Apriori_1$. The proposed algorithm is as follows:

T: Transactional data base
$C_k$: Candidate item set of size k
$L_k$: Frequent item set of size k
s: Support

$Apriori_2$(T, s)
    $L_1 \leftarrow$ { large 1-item set that appear in more than or equal to s transactions }
    $k \leftarrow 2$
    While $L_{k-1} \neq \phi$
      $C_k \leftarrow Join(L_{k-1})$
      For each candidate c in $C_k$
        $L_k = \phi$
        $T_{id} = Get\_Common\_Transaction\_Ids(c, L_1)$
        If ( $|T_{id}| >= s$) then
          $L_k \leftarrow L_k \cup \{c\}$
        End If
      End For
    $k \leftarrow k + 1$
    End While
    Return $L_k$
End $Apriori_2$

The proposed algorithm $Apriori_2$ combines both join and prune operations of Apriori and $Apriori_1$. The proposed algorithm obtains common transaction ids for all items in c of $C_k$ as a group ($T_{id}$). The count of group of transaction ids ($|T_{id}|$) defines the frequency of c. If this frequency is greater than and equal to support, s then c of $C_k$ is included in $L_k$. The number of statements in proposed algorithm $Apriori_2$ is less when compared with Apriori and $Apriori_1$.

4.2 Example

To trace out the proposed algorithm consider the transactional database with transactions (Ti) and items ($I_i$) shown in Table 1. This considered transactional data set consists of nine transactions and five items. Assume the value of support, s as 3.

Table 1: Transactional Database

| $T_{id}$ | Items |
|---|---|
| $T_1$ | $I_1, I_2, I_5$ |
| $T_2$ | $I_2, I_4$ |
| $T_3$ | $I_2, I_4$ |
| $T_4$ | $I_1, I_2, I_4$ |
| $T_5$ | $I_1, I_3$ |
| $T_6$ | $I_2, I_3$ |
| $T_7$ | $I_1, I_3$ |
| $T_8$ | $I_1, I_2, I_3, I_5$ |
| $T_9$ | $I_1, I_2, I_3$ |

At first the proposed algorithm generates candidate 1-item set. This is shown in Table 2.

Table 2: Candidate 1 – item set

| Item | Frequency |
|---|---|
| $I_1$ | 6 |
| $I_2$ | 7 |
| $I_3$ | 5 |
| $I_4$ | 3 |
| $I_5$ | 2 |

The items with frequency greater than or equal to 3 are considered and remaining items are excluded. This results in generation of frequent 1 – item set, $L_1$. This is shown in Table 3. $I_5$ is excluded because its frequency is less than 3.

Table 3: Frequent 1 – item set ($L_1$)

| Item | Frequency | Transaction Ids |
|---|---|---|
| $I_1$ | 6 | $T_1, T_4, T_5, T_7, T_8, T_9$ |
| $I_2$ | 7 | $T_1, T_2, T_3, T_4, T_6, T_8, T_9$ |
| $I_3$ | 5 | $T_5, T_6, T_7, T_8, T_9$ |
| $I_4$ | 3 | $T_2, T_3, T_4$ |

The frequent 2 – item set is generated using common transaction ids. This is shown in Table 4.

Table 4: Frequent 2-item set ($L_2$)

| Item | Common Transaction Ids | Count of common transaction ids ($|T_{id}|$) |
|---|---|---|
| $I_1, I_2$ | $T_1, T_4, T_8, T_9$ | 4 |
| $I_1, I_3$ | $T_5, T_7, T_8, T_9$ | 4 |
| $I_2, I_3$ | $T_6, T_8, T_9$ | 3 |
| $I_2, I_4$ | $T_2, T_3, T_4$ | 3 |

Consider the candidate ($I_1, I_2$). To calculate frequency of ($I_1, I_2$) the general Apriori involves in scanning all transactions (9 transactions) and $Apriori_1$ involves in scanning six transactions ($T_1, T_4, T_5, T_7, T_8, T_9$). The proposed algorithm involves in selecting the common transaction ids which contains ($I_1, I_2$) which doesn't include of scanning the transactions for checking the presence of ($I_1, I_2$) because obtaining the common transaction ids containing the candidate ($I_1, I_2$) need not scan transactions. ($I_1, I_4$) and ($I_3, I_4$) are not included into $L_2$ because their $|T_{id}|$ is less than 3. Frequent 3 – item set is empty because the count of common transaction ids for all candidates in $C_3$ is less than 3.

## 5. Analysis of Proposed Algorithm $Apriori_2$

The number of scans by Apriori, $Apriori_1$ and $Apriori_2$ over the transactional database shown in Table 1 are tabulated is Table 5. This is tabulated with support (s) as 3

Table 5: Number of scans for generations of L1, L2 and $L_3$ by Apriori, $Apriori_1$ and $Apriori_2$

|  | Apriori | $Apriori_1$ | $Apriori_2$ |
|---|---|---|---|
| Frequent 1-item set | 45 | 45 | 45 |
| Frequent 2-item set | 54 | 25 | 0 |
| Frequent 3-item set | 36 | 14 | 0 |
| Total no. of scans | 135 | 84 | 45 |

The number of scans doesn't alter by $Apriori_2$ even when the support (s) value is increased whereas the number of scans for Apriori and $Apriori_1$ alters.

The time consumed for generating frequent items ($L_k$) in milliseconds by general Apriori algorithm, $Apriori_1$ and $Apriori_2$ is shown in table 6. This is calculated over the transactional database shown in Table 1 with support (s) as 3.

Table 6: Time (in milli seconds) for generation of $L_2$ and $L_3$ by Apriori, $Apriori_1$ and $Apriori_2$

| Time in milliseconds | | | |
|---|---|---|---|
|  | Apriori | $Apriori_1$ | $Apriori_2$ |
| Frequent 2-item set | 14 | 11 | 9 |
| Frequent 3-item set | 15 | 13 | 12 |

The execution time decreases with increase in value of support(s).

## 6. Conclusions

The proposed algorithm $Apriori_2$, reduces number of scans of transactional database while generating $L_k$. Execution time also improved when compared with Apriori and $Apriori_1$. This represents that $Apriori_2$ takes less time in generating frequent patterns. If the value of support is increased then the number of scans also gets decreased (for generating $L_1$ number of scans decreases and for $L_2$ and above the number of scans always remains zero). The proposed algorithm fits for larger transactional databases because the proposed algorithm doesn't scan transactional database while generating $L_k$ (k>1). It looks into only transactional ids. This proposed algorithm is apt for problems in areas like marketing, whether, health care etc.

## References

[1] J. Han, and M. Kamber, "Data Mining: Concepts and Techniques", Morgan Kaufmann Publishers, 2000.
[2] J. D. Holt, and S. M. Chung, "Efficient Mining of Association Rules in Text Databases", in CIKM'99, Nov 1999, Kansas City, USA, pp. 234242.
[3] B. Mobasher, N. Jain, E.H. Han, and J. Srivastava, "Web Mining: Pattern Discovery from World Wide Web Transactions", Department of Computer Science, University of Minnesota, March 1996, Technical Report TR96-050.
[4] C. Ordonez, and E. Omiecinski, "Discovering Association Rules Based on Image Content", IEEE Advances in Digital Libraries, 1999.
[5] R. Agrawal, T. Imielinski, and A. Swami, "Mining association rules between sets of items in large databases", in ACM SIGMOD International Conference on Management of Data, May 1993, Washington, USA, pp. 207216.
[6] R. Agrawal, H. Mannila, R. Srikant, H. Toivonen, and A. I. Verkamo, "Fast discovery of association rules. In Advances in Knowledge Discovery and Data Mining", AAAI Press, 1996, pp. 307328.
[7] R. Bayardo, and R. Agrawal, "Mining the most interesting rules", in 5th International Conference on Knowledge Discovery and Data Mining, San Diego, August 1999, California, USA, pp. 145154.
[8] J. Hipp, U. Güntzer, and U. Grimmer, "Integrating association rule mining algorithms with relational database

systems", in 3rd International Conference on Enterprise Information Systems, July 2001, Setúbal, Portugal, pp. 130137.
[9] R. Ng, L. S. Lakshmanan, J. Han, and T. Mah, "Exploratory mining via constrained frequent set queries", in ACM-SIGMOD International Conference on Management of Data, June 1999, Philadelphia, PA, USA, pp. 556558.
[10] Y. Guizhen, "The complexity of mining maximal frequent itemsets and maximal frequent patterns", in ACM SIGKDD International Conference on Knowledge Discovery and Data Mining , August 2004, Seattle, WA, USA, pp. 343353.
[11] L. Klemetinen, H. Mannila, P. Ronkainen, et al., "Finding interesting rules from large sets of discovered association rules", in Third International Conference on Information and Knowledge Management, 1994, Gaithersburg, USA, pp. 401407.
[12] J. S. Park, M.S. Chen, and P.S. Yu, "An Effective HashBased Algorithm for Mining Association Rules", in ACM SIGMOD International Conference on Management of Data, 1995, San Jose, CA, USA, pp. 175186.
[13] H. Toivonen, "Sampling large databases for association rules", in 22nd International Conference on Very Large Data Bases, 1996, pp. 134–145.
[14] P. Kotásek, and J. Zendulka, "Comparison of Three Mining Algorithms for Association Rules", in 34th Spring International Conference on Modelling and Simulation of Systems, 2000, Workshop Proceedings Information Systems Modelling, pp. 8590.
[15] J. Han, J. Pei, and Y. Yin, "Mining frequent patterns Candidate generation", in ACMSIGMOD International Management of Data, 2000, Dallas, TX.
[16] F. H. AL-Zawaidah, Y. H. Jbara, and A. L. Marwan, "An Improved Algorithm for Mining Association Rules in Large Databases", International Journal on Natural Language Computing , Vol. 1, No. 7, 2011, pp. 311-316.
[17] J. Han, M. Kamber, "Data Mining: Concepts and Techniques", *Morgan Kaufmann Publishers*, Book, 2000.
[18] Mohammed Al-Maolegi1, and Bassam Arkok2, "International Journal on Natural Language Computing", Vol. 3, No.1, February 2014.

**Sudhir Tirumalasetty,** Associate Professor in Department of Computer Science & Engineering, Vasireddy Venkatadri Institute of Technology. Pursuing Ph.D in Computer Science & Engineering from Acharya Nagarjuna University, Guntur, Andhra Pradesh, India. Area of research is Data Mining.

**Aruna Jadda,** pursuing M.Tech in Computer Science & Engineering from Vasireddy Venkatadri Institute of Technology affiliated to Jawaharlal Nehru Technological University, Kakinada, Andhra Pradesh, India

**Dr. Sreenivasa Reddy Edara,** Dean for University College of Engineering, Acharya Nagarjuna University, Guntur, Andhra Pradesh, India. Done Ph.D in Computer Science & Engineering from Acharya Nagarjuna University.